%Paper: hep-th/9503071
%From: "Serge E. Parkhomenko" <spark@itp.ac.ru>
%Date: Fri, 10 Mar 1995 20:02:41 +0300

% PLAINTEX FILE, NO MACROS
%
\magnification=\magstep1
\def\lgl{\langle}
\def\rgl{\rangle}
\def\rbr{\rbrack}
\def\lbr{\lbrack}
\def\rbrc{\rbrace}
\def\lbrc{\lbrace}
\def\ov{\over}
\def\d{\partial}

%\begin{document}

%\nopagenumbers
\rightline{Landau Tmp/12/93.}
\rightline{March, 1995}

\vskip 2 true cm
\centerline{Quasi Frobenius Lie algebras construction
of N=4 superconformal field theories.}
\vskip 2.5 true cm
\centerline{S. E. Parkhomenko\footnote{$\sp{\dag}$}
 {e-mail address:spark@itp.chg.free.net}}
\vskip 0.5 true cm
\centerline{\vbox{\hbox{Landau Institute for Theoretical Physics}
                  \hbox{\sevenrm{142432 Chernogolovka,Russia}}
                  \hbox{}
                  \hbox{}}}
\vskip 1 true cm
\centerline{\bf Abstract}
\vskip 0.5 true cm
 Manin triples construction of N=4 superconformal field theories
is considered. The correspondence between quasi Frobenius
finite-dimensional Lie algebras and N=4 Virasoro superalgebras
is established.
\smallskip
\vskip 10pt
\centerline{\bf Introduction.}
 In connection with numerous string applications extended
superconformal field theories (SCFT's) in two dimensions have
became increasingly important over the past few years.
It is know that a large class of extended SCFT's is obtainable
from Kazama-Suzuki models [1], that is, the supercoset constructions
assochiated with compact Kahler homogeneous spaces $G/H$.
In [2-3] WZNW models were studed, which allow for extended
supersymmetry and conditions were formulated that the Lie group
must satisfy so that its WZNW model would have extended supersymmetry.
In particular, in [3] a correspondence was established between
$N=2, 4$ SCFT's and finite-dimensional Manin triples. From the point
of view accepted in [3] Kazama-Suzuki models are particular cases
of Manin triple construction of extended SCFT. Indeed they correspond
to Manin triples associated with any simple Lie algebra and its
parabolic subalgebra. It is intresting to note that there is
a similar construction of $N=3/2$ SCFT based on finite-dimensional
Manin pairs [4].
 In this paper the conditions under which $N=2$ SCFT admit
$N=4$ supersymmetric extensions will be investigated from
more general positions than it was done in [2-3].
 The paper is arranged as follows. In Section 2 we breafly
review the Manin triples construction of $N=2$ SCFT's.
In Section 3 we investigate the conditions under which
a $N=2$ SCFT associated with any finite-dimensional
Manin triple possess $N=4$ Virasoro superalgebra of symmetries.
We will see that it is possible to construct generators
of $N=4$ Virasoro superalgebra if the isotropic
subalgebras of Manin triple are quasi Frobenius Lie algebras.
Moreover if they are Frobenius then it is
possible to construct generators of two different
$N=4$ Virasoro superalgebras. This case corresponds to
the "big" $N=4$ Virasoro superalgebra constructed in [2]
and investigated in [7-9].
In section 4 we give some examples of our construction.

%%%%%%%%%%%%%%%%%%%%%%%%%%%%%%%%%%%%%%%%%%%%%%%%%%%%%
\vskip 10pt
\centerline{\bf2. N=2 SCFT and finite-dimensional Manin triples.}

 We begin with the definition of Manin triple [5]

DEFINITION 2.1. A Manin triple $(g,g_{+},g_{-})$ consists of
a Lie algebra $g$, with nondegenerate invariant inner product
$(,)$ and isotropic Lie subalgebras $g_{\pm}$ such that
$g=g_{+}\oplus g_{-}$ as vector space.

For any finite-dimensional Manin triple let us fix any
orthonormal basis $\{E^{a}, E_{a}, a=1,...,d\}$ in algebra
$g$ so that $\{E^{a}\}$- basis in $g_{+}$, $\{E_{a}\}$- basis
in $g_{-}$. The brakets and Jacoby identity of $g$
are given by
$$
\eqalign { \lbr E^{a},E^{b}\rbr=f^{ab}_{c}E^{c}     \cr
           \lbr E_{a},E_{b}\rbr=f_{ab}^{c}E^{c}     \cr
           \lbr E^{a},E_{b}\rbr=f_{bc}^{a}E^{c}-f^{ac}_{b}E_{c} }\eqno(2.1)
$$
$$
\eqalign{ f^{ab}_{d}f^{dc}_{e}+f^{bc}_{d}f^{da}_{e}+
          f^{ca}_{d}f^{db}_{e}=0                                          \cr
          f_{ab}^{d}f_{dc}^{e}+f_{bc}^{d}f_{da}^{e}+
          f_{ca}^{d}f_{db}^{e}=0                                          \cr
          f_{mc}^{a}f^{bm}_{d}-f_{md}^{a}f^{bm}_{c}-f_{mc}^{b}f^{am}_{d}+ \cr
          f_{md}^{b}f^{am}_{c}= f_{cd}^{m}f^{ab}_{m} }\eqno(2.2)
$$
In the following we will be needed in the consequence of (2.2)
$$
 f_{m}f^{mb}_{a}+f^{m}f_{ma}^{b}=-f_{nm}^{b}f^{mn}_{a}  \eqno(2.3)
$$
, where $f_{m}=f_{ma}^{a}$, $f^{m}=f^{ma}_{a}$.
Denote by $\langle,\rangle$ the Killing form of $g$. It is not difficult
to calculate
$$
\eqalign {\langle E^{a}, E^{b}\rangle= 2f^{ac}_{d}f^{bd}_{c} \cr
          \langle E_{a}, E_{b}\rangle= 2f_{ac}^{d}f_{bd}^{c} \cr
          \langle E^{a}, E_{b}\rangle= -f^{cd}_{b}f_{cd}^{a}-
          2f^{ac}_{d}f_{bc}^{d} }\eqno(2.4)
$$
Let us denote
$$
\eqalign {B^{b}_{a}= f_{c}f^{cb}_{a}+f^{c}f_{ca}^{b} \cr
          A^{b}_{a}= f_{ac}^{d}f^{bc}_{d} }\eqno(2.5)
$$
Then we will have
$$
 \langle E^{b},E_{a}\rangle= -B^{b}_{a}-2A^{b}_{a} \eqno(2.6)
$$
Let $J^{a}(z), J_{a}(z)$ be the generators of affine Kac-Moody
algebra $\hat{g}$, which correspond to the fixed
basis $\{E^{a}, E_{a}\}$, so that currents $J^{a}$ generate
subalgebra $\hat{g_{+}}$ and currents $J_{a}$ generate
subalgebra $\hat{g_{-}}$. The singular OPE's between these
currents is the following
$$
\eqalign{ J^{a}(z)J^{b}(w)=-(z-w)^{-2}{1\ov 2}\lgl E^{a},E^{b}\rgl
          +(z-w)^{-1}f^{ab}_{c}J^{c}(w)+reg   \cr
          J_{a}(z)J_{b}(w)=-(z-w)^{-2}{1\ov 2}\lgl E_{a},E_{b}\rgl
          +(z-w)^{-1}f_{ab}^{c}J_{c}(w)+reg   \cr
          J^{a}(z)J_{b}(w)=(z-w)^{-2}{1\over 2}
          (q\delta^{a}_{b}-\lgl E^{a},E_{b}\rgl)+             \cr
          (z-w)^{-1}(f_{bc}^{a}J^{c}-f^{ac}_{b}J_{c})(w)+reg }\eqno(2.7)
$$
,where $q=2(k+v)$, $v= {1\over 2d}\sum_{i}Tr(adE^{i}adE^{i})$
and $E^{i}=E^{a}, i=a, E^{i}=E_{a}, i=a+d$.
Let $\psi^{a}(z), \psi_{a}(z)$ be free fermion currents which
have singular OPE's with respect to the inner product $(,)$
$$
 \psi^{a}(z)\psi_{b}(w)= (z-w)^{-1}\delta^{a}_{b}+reg \eqno(2.8)
$$

ASSERTION 2.2 [3,4] The currents
$$
\eqalign{ G^{+}=\sqrt{{2\over k+v}}(\psi^{a}J_{a}-
          {1\over 2}f_{ab}^{c}\psi^{a}\psi^{b}\psi_{c})   \cr
          G^{-}=\sqrt{{2\over k+v}}(\psi_{a}J^{a}-
          {1\over 2}f^{ab}_{c}\psi_{a}\psi_{b}\psi^{c})   \cr
          K= (\delta^{b}_{a}+{B^{b}_{a}\over k+v}):\psi^{a}\psi_{b}:+
          {1\over k+v}(f_{c}J^{c}-f^{c}J_{c})             \cr
          2T={1\over k+v}:(J^{a}J_{a}+J_{a}J^{a}):+       \cr
          :(\d \psi^{a}\psi_{a}-\psi^{a}\d \psi_{a}): }\eqno(2.9)
$$
satisfy the operator products of the $N=2$ Virasoro superalgebra
with central charge
$$
 c= 3({D\over2}+{A^{a}_{a}\over k+v}) \eqno(2.10)
$$

DEFINITION 2.3. Let $g$ be the Lie algebra with nondegenerate
invariant inner product $(,)$ and $R$- complex structure
on vector space $g$ skew- symmetric relative to $(,)$.
$R$ is complex stucture on Lie algebra $g$ if $R$ satisfies
the modified classical Yang-Baxter equation:
$$
 [Rx, Ry]- R[Rx, y]- R[x, Ry]= [x, y]  \eqno(2.11)
$$

 It is not dificult to establish the correspondence between
complex Manin triples and complex structures on Lie algebras [3].
Namely for any complex Manin triple $(g,g_{+},g_{-})$
there is canonic complex structure on Lie algebra such
that subalgebras $g_{\pm}$ are $\pm \imath$- ei\-gen\-spa\-ces
of its. On the other hand, for any real Lie algebra $g$
with nondegenerate invariant inner product and skew-
symmetric complex structure $R$ on this algebra one can
consider the complexification $g_{C}$ of $g$. Let
$g_{\pm}$ be $\pm \imath$- eigenspaces of $R$
in algebra $g_{C}$, then $(g_{C},g_{+},g_{-})$ be the
complex Manin triple. Hence we can use formulas (2.9)
to build up generators of $N=2$ Virasoro superalgebra.

In connection with the construction described above
it is pertient to note the work [13] where very
similar construction was considered.

%%%%%%%%%%%%%%%%%%%%%%%%%%%%%%%%%%%%%%%%%%%%%%%%%%%%%%%%%%%%%%%%%%%%%

\vskip 10pt
\centerline{\bf3. Quasi Frobenius Lie algebras and N=4
                  Virasoro superalgebras.}

Now we will try to generalize Manin triples construction
of $N=$ SCFT for $N=4$ SCFT. $N=4$ Virasoro superalgebra
have the following OPE
$$
\eqalign{ T(z)T(w)=(z-w)^{-4}{c\over 2}+(z-w)^{-2}2T(w)+
          (z-w)^{-1}\d T(w)+reg \cr
          K^{i}(z)K^{j}(w)=(z-w)^{-2}{c\over 12}+
          (z-w)^{-1}\imath \varepsilon^{ijk}K^{k}(w)+reg \cr
          T(z)K^{i}(w)=(z-w)^{-2}K^{i}(w)+
          (z-w)^{-1}\d K^{i}(w)+reg \cr
          K^{i}(z)G^{a}(w)=
          -(z-w)^{-1}{1\over2}(\sigma^{i})^{a}_{b}G^{b}(w)+reg \cr
          K^{i}(z)G_{a}(w)=
          (z-w)^{-1}{1\over2}(\sigma^{i})^{b}_{a}G_{b}(w)+reg \cr
          T(z)G^{a}(w)=(z-w)^{-2}{3\over 2}G^{a}(w)+
          (z-w)^{-1}\d G^{a}(w)+reg \cr
          T(z)G_{a}(w)=(z-w)^{-2}{3\over 2}G_{a}(w)+
          (z-w)^{-1}\d G_{a}(w)+reg \cr
          G^{a}(z)G_{b}(w)=(z-w)^{-3}{2c\over 3}\delta^{a}_{b}
          +(z-w)^{-2}4(\sigma^{i})^{a}_{b}K^{i}(w)+ \cr
          (z-w)^{-1}(2\delta^{a}_{b}T(w)+
          2(\sigma^{i})^{a}_{b}\d K^{i}(w))+reg }\eqno(3.1)
$$
Let us fix some finite- dimensional Manin triple $(g,g_{+},g_{-})$.
{}From the formulas (3.1) we can see that currents $G^{0}, G_{0}$
generate $N=2$ Virasoro superalgebra. With the arguments of
preceding section one can establish the existence of the complex
structure $R_{1}$ on Lie algebra $g$. Let us denote:
$$
\eqalign{ D_{0}= {1\ov \sqrt{2}}(G^{0}+G_{0}) \cr
          D_{1}= {-\imath \ov \sqrt{2}}(G^{0}-G_{0}) \cr
          D_{2}= {1\ov \sqrt{2}}(G^{1}+G_{1}) }\eqno(3.2)
$$
We can see from (3.1) that the linear combinations
$$
 {1\ov \sqrt{2}}(D_{0}\pm \imath D_{2})=
 G^{0}+G_{0}\pm \imath(G^{1}+G_{1}) \eqno(3.3)
$$
generate another $N=2$ Virasoro superalgebra.
Therefore we establish the existence of the second
complex stucture $R_{2}$ on the Lie algebra $g$.
Using (3.1) once more it is not dificult to show
that for any real numbers $x$ and $y$, such that
$x^{2}+ y^{2}=1$ the currents
$$
 {1\ov \sqrt{2}}(D_{0}\pm \imath (xD_{1}+ yD_{2}))
$$
also generate $N=2$ Virasoro superalgebra. Hence with the
arguments of preceding section we conclude, that the
square of the operator
$$
 S= xR_{1}+ yR_{2} \eqno(3.4)
$$
is equal to $-1$. This fact implies
$$
 R_{1}R_{2}+ R_{2}R_{1}=0 \eqno(3.5)
$$
Next, we intend to show that the existence of two skew-
symmetric mutualy anticommuting complex structures
on Lie algebra makes it possible to construct generators
of $N=4$ Virasoro superalgebra.

 Let $g_{\pm}$ be the eigenspaces of the complex structure
$R_{1}$ on complex Lie algebra $g$. Let us fix the
orthonormal basis (2.1) in $g$. In this basis
the second complex structure $R_{2}$ is given by matrix
$$
\eqalign{ R_{2}E^{a}= (r_{11})^{a}_{b}E^{b}+(r_{12})^{ab}E_{b} \cr
          R_{2}E_{a}= (r_{22})^{b}_{a}E_{b}+(r_{21})_{ab}E^{b} }\eqno(3.6)
$$
and the equation (3.5) equivalet to
$$
 r_{11}= r_{22}= 0 \eqno(3.7)
$$
The skew- symmetric condition for $R_{2}$
takes the form
$$
 r_{12}^{T}=-r_{12},\ r_{21}^{T}=-r_{21}  \eqno(3.8)
$$
Taking into account (3.7), (3.8) one can rewrite
the equation $(R_{2})^{2}= -1$ in the following form
$$
 r_{21}= -r_{12}^{-1} \eqno(3.9)
$$
In the following we will denote $r_{12}$ as $r$.
In the basis (2.1) equation (2.11) for the $R_{2}$
takes the form
$$
\eqalign{ r_{ad}f_{cb}^{d}+r_{bd}f_{ac}^{d}+r_{cd}f_{ba}^{d}= 0 \cr
          r^{ad}f^{cb}_{d}+r^{bd}f^{ac}_{d}+r^{cd}f^{ba}_{d}= 0 }
          \eqno(3.10)
$$
,where $r^{ab}=(r^{-1})_{ab}$.
That is $r$ is 2-cocycle on algebra $g_{+}$
and $r^{-1}$ is 2-cocycle on $g_{-}$. In view of (3.9)
they should be nondegenerate.

 Define fermionic currents
$$
\eqalign{ G^{0}_{x}= \psi^{a}J_{a}-
          {1\over 2}f_{ab}^{c}\psi^{a}\psi^{b}\psi_{c}+
          x^{0}_{a}\d \psi^{a}                                            \cr
          G_{0x}= \psi_{a}J^{a}-
          {1\over 2}f^{ab}_{c}\psi_{a}\psi_{b}\psi^{c}+
          x_{0}^{a}\d \psi_{a}                                            \cr
          G^{1}_{x}= r_{ba}\psi^{a}J^{b}+
          {1\over 2}r_{am}f^{ab}_{c}r_{bn}r^{ck}
          \psi^{m}\psi^{n}\psi_{k}+
          x^{1}_{a}\d \psi^{a}                                \cr
          G_{1x}= r^{ba}\psi_{a}J_{b}+
          {1\over 2}r^{am}f_{ab}^{c}r^{bn}r_{ck}
          \psi_{m}\psi_{n}\psi^{k}+
          x_{1}^{a}\d \psi_{a} }\eqno(3.11)
$$
,where vectors $x^{0}, x_{0}, x^{1}, x_{1}$ will be
determined later and denote
$$
\eqalign{ G^{0}_{x}= G^{0}+\d x^{0},\quad
          G_{0x}= G_{0}+\d x_{0} \cr
          G^{1}_{x}= G^{1}+\d x^{1},\quad
          G_{1x}= G_{1}+\d x_{1} }\eqno(3.12)
$$
It is not difficult to show that the conditions
that there no singular terms in OPE's $G^{0}_{x}G^{0}_{x}$
, $G_{0x}G_{0x}$, $G^{1}_{x}G^{1}_{x}$, $G_{1x}G_{1x}$
are equivalent to the equations
$$
\eqalign{ x^{0}_{a}f_{bc}^{a}=0,\quad
          x_{0}^{a}f^{bc}_{a}=0 \cr
          x^{1}_{a}r^{ad}f^{bc}_{d}=0,\quad
          x_{1}^{a}r_{ad}f_{bc}^{d}=0 }\eqno(3.13)
$$
(here we have used (3.10)), that is vectors
$x^{0}, x_{0}, rx^{1}, r^{-1}x_{1}$ are 1-cocycles on
subalgebras $g_{\pm}$. From now we will imply that (3.13) be
satisfied.

LEMMA 3.1.
$$
\eqalign{ G^{0}_{x}(z)G^{1}_{x}(w)= (z-w)^{-2}\lbr
          {q\ov2}r_{ac}-f_{ac}^{m}f^{n}r_{mn}                        \cr
          -{1\ov2}x^{1}_{m}f_{ac}^{m}
          -{1\ov2}x^{0}_{b}(r_{pa}f^{pb}_{c}-r_{pc}f^{pb}_{a})
          \rbr \psi^{a}\psi^{c}+                                     \cr
          (z-w)^{-1}{1\ov2}\lbr
          {q\ov2}r_{ac}-f_{ac}^{m}f^{n}r_{mn}
          -{1\ov2}x^{1}_{m}f_{ac}^{m}\rbr \d (\psi^{a}\psi^{c})+reg. }
          \eqno(3.14a)
$$
$$
\eqalign{ G^{0}_{x}(z)G_{1x}(w)=
          -(z-w)^{-3}2x^{0}_{a}x^{a}_{1}                             \cr
          -(z-w)^{-2}2\lbr (f_{a}r^{ab}-{x^{b}_{1}\ov2})(J_{b}+
          f_{bd}^{c}\psi_{c}\psi^{d})+
          {x^{0}_{a}\ov2}r^{ca}(J_{c}+
          f_{cb}^{d}r^{bn}r_{dk}\psi_{n}\psi^{k})\rbr                \cr
          -(z-w)^{-1}(f_{a}r^{ab}-x^{b}_{1})(J_{b}+
          f_{bd}^{c})\psi_{c}\psi^{d}+reg. }
          \eqno(3.14b)
$$
$$
\eqalign{ G_{0x}(z)G^{1}_{x}(w)=
          -(z-w)^{-3}2x_{0}^{a}x_{a}^{1}                             \cr
          -(z-w)^{-2}2\lbr (f^{a}r_{ab}-{x^{1}_{b}\ov2})(J^{b}+
          f^{bd}_{c}\psi^{c}\psi_{d})+
          {x_{0}^{a}\ov2}r_{ca}(J^{c}+
          f^{cb}_{d}r_{bn}r^{dk}\psi^{n}\psi_{k})\rbr                \cr
          -(z-w)^{-1}(f^{a}r_{ab}-x_{1}^{b})(J^{b}+
          f^{bd}_{c}\psi^{c}\psi_{d})+reg. }
          \eqno(3.14c)
$$
$$
\eqalign{ G_{0x}(z)G_{1x}(w)= (z-w)^{-2}\lbr
          {q\ov2}r^{ac}-f^{ac}_{m}f_{n}r^{mn}                        \cr
          -{1\ov2}x_{1}^{m}f^{ac}_{m}
          -{1\ov2}x_{0}^{b}(r^{pa}f_{pb}^{c}-r^{pc}f_{pb}^{a})
          \rbr \psi_{a}\psi_{c}+                                      \cr
          (z-w)^{-1}{1\ov2}\lbr
          {q\ov2}r^{ac}-f^{ac}_{m}f_{n}r^{mn}
          -{1\ov2}x_{1}^{m}f^{ac}_{m}\rbr \d (\psi_{a}\psi_{c})+reg. }
          \eqno(3.14d)
$$

PROOF.
Let us calculate the operator product
$$
\eqalign{ G^{0}_{x}(z)G^{1}_{x}(w)=G^{0}(z)G^{1}(w)+G^{0}(z)\d x^{1}(w)
          +\d x^{0}(z)G^{1}(w)+\d x^{0}(z)\d x^{1}(w) }\eqno(3.15)
$$
We start by calculating singular terms of $G^{0}G^{1}$
$$
\eqalign{ G^{0}(z)G^{1}(w)= (z-w)^{-2}\lbr
 {1\ov2}r_{bc}(q\delta^{b}_{a}-<E^{b},E_{a}>)\psi^{a}\psi^{c}+   \cr
 f_{am}^{n}f^{bc}_{d}r_{bn}r_{cq}r^{dm}\psi^{a}\psi^{q}\rbr+     \cr
 (z-w)^{-1}\lbr
 {1\ov2}r_{bc}(q\delta^{b}_{a}-<E^{b},E_{a}>)\d \psi^{a}\psi^{c}+\cr
 f_{am}^{n}f^{bc}_{d}r_{bn}r_{cq}r^{dm}\d \psi^{a}\psi^{q}\rbr+  \cr
 (z-w)^{-1}{1\ov2}\lbr
 -f_{am}^{n}f^{bc}_{d}r_{bn}r_{cq}r^{ds}\psi^{a}\psi^{m}\psi^{q}\psi_{s}+\cr
 f_{am}^{n}f^{bc}_{d}r_{bp}r_{cq}r^{dm}\psi^{a}\psi^{p}\psi^{q}\psi_{n}\rbr }
 \eqno(3.16)
$$
Let us denote
$$
\eqalign{
 U={1\ov2}r_{bc}(q\delta^{b}_{a}-<E^{b},E_{a}>)\psi^{a}\psi^{c}+ \cr
 f_{am}^{n}f^{bc}_{d}r_{bn}r_{cq}r^{dm}\psi^{a}\psi^{q}\rbr   \cr
 V={1\ov2}r_{bc}(q\delta^{b}_{a}-<E^{b},E_{a}>)\d \psi^{a}\psi^{c}+\cr
 f_{am}^{n}f^{bc}_{d}r_{bn}r_{cq}r^{dm}\d \psi^{a}\psi^{q}\rbr  \cr
 W={1\ov2}(
 -f_{am}^{n}f^{bc}_{d}r_{bn}r_{cq}r^{ds}\psi^{a}\psi^{m}\psi^{q}\psi_{s}+\cr
 f_{am}^{n}f^{bc}_{d}r_{bp}r_{cq}r^{dm}\psi^{a}\psi^{p}\psi^{q}\psi_{n}) }
$$
We are going to show that
$$
 W=0  \eqno(3.17)
$$
In view of (3.9), (3.10) we have
$$
\eqalign{
 f^{bc}_{d}r_{bn}r_{cq}r^{ds}= f^{bs}_{q}r_{bn}-f^{bs}_{n}r_{bq} \cr
 f_{bc}^{d}r^{bp}r^{cq}r_{dm}= f_{bm}^{q}r^{bp}-f_{bp}^{n}r^{bq} }
 \eqno(3.18)
$$
Taking this equation into account we can represent $W$ as
the following
$$
\eqalign{
  W= ((f^{it}_{k}r_{iq}-f^{it}_{q}r_{ik})f_{ap}^{k}+ \cr
  (f^{im}_{a}r_{ip}-f^{im}_{q}r_{ia})f_{mq}^{k})
  \psi^{a}\psi^{p}\psi^{q}\psi_{t} }\eqno(3.19)
$$
Here it is pertient to make a comment about (3.17).
Let us denote
$$
 h_{qn}^{s}=f^{bs}_{q}r_{bn}-f^{bs}_{n}r_{bq} \eqno(3.20)
$$
In view of (3.9), (3.10) the constants $h_{qn}^{s}$
determine another Lie structure on vector space $g_{-}$.
{}From (3.19) we can see that (3.17) is the
condition this new Lie- structure is compatible with
the old Lie- structure on $g_{-}$.
Using the first equation from (3.10) one can write
$$
\eqalign{
 (f^{it}_{k}r_{iq}-f^{it}_{q}r_{ik})f_{ap}^{k}= f^{it}_{k}f_{ap}^{k}r_{iq}+\cr
 f^{it}_{q}(r_{ak}f_{pi}^{k}+r_{pk}f_{ia}^{k}) }\eqno(3.21)
$$
After sbstitution this formula in the left hand side
of (3.19) we obtain
$$
\eqalign{
 W= (f_{ap}^{i}f^{kt}_{i}r_{kq}+2f_{pi}^{k}f^{it}_{a}r_{kq}-
 2f_{pi}^{t}f^{ik}_{a}r_{kq})\psi^{a}\psi^{p}\psi^{q}\psi_{t} }\eqno(3.22)
$$
Now (3.17) follows from the last equation out of (2.2).
With help of (3.10) and (2.6) $U$ can be transformed
into
$$
 U=({q\ov 2}r_{ac}+{1\ov 2}B^{b}_{a}r_{bc}-
 {1\ov 4}f_{ac}^{m}f^{nb}_{m}r_{bn})\psi^{a}\psi^{c} \eqno(3.23)
$$
Using (2.5), (3.10) it is not difficult to obtain
$$
 B^{b}_{a}r_{bc}\psi^{a}\psi^{c}=
 -{1\ov 2}f_{ca}^{m}f^{nk}_{m}r_{nk}\psi^{a}\psi^{c} \eqno(3.24)
$$
{}From (3.10) we can easily find
$$
\eqalign{ f^{nk}_{m}r_{nk}= 2f^{n}r_{nm} \cr
          f_{nk}^{m}r^{nk}= 2f_{n}r^{nm} }\eqno(3.25)
$$
Hence, we have
$$
 U=({q\ov 2}r_{ab}-f_{ab}^{m}f^{n}r_{mn})\psi^{a}\psi^{c} \eqno(3.26)
$$
Next, transform V
$$
\eqalign{
 V= {1\ov 2}({1\ov 2}(q\delta_{a}^{b}+B^{b}_{a}+2A_{a}^{b})r_{bc}+
 f_{am}^{n}f^{pq}_{s}r_{pn}r_{qc}r^{sm})\d (\psi^{a}\psi^{c})+ \cr
 {1\ov 2}({1\ov 2}q\delta_{a}^{b}+B_{a}^{b}+2A_{a}^{b})r_{bc}+
 f_{am}^{n}f^{pq}_{s}r_{pn}r_{qc}r^{sm})
 (\d \psi^{a}\psi^{c}-\psi^{a}\d \psi^{c}) }\eqno(3.27)
$$
{}From (3.18), (2.5) one can get
$$
 f_{am}^{n}f^{ps}_{t}r_{pn}r_{sc}r^{tm}=
 -{1\ov 2}B_{c}^{b}r_{ba}-A_{a}^{b}r_{bc}  \eqno(3.28)
$$
Therefore the first term from this expression
is equal to ${1\ov  2}\d U$ and the second term
is equal to zero.
%From (3.18) one can get
%$$
%\eqalign{
% f_{am}^{n}f^{pq}_{s}r_{pn}r_{qc}r^{sm}+
% f_{cm}^{n}f^{pq}_{s}r_{pn}r_{qa}r^{sm}= \cr
% r_{kn}f_{am}^{n}f^{km}_{c}+r_{kn}f_{cm}^{n}f^{km}_{a}-
% r_{kc}A^{k}_{a}-r_{ka}A^{k}_{c} }
% \eqno(3.30)
%$$
%From (2.3), (2.5), (3.10) one can obtain
%$$
% 2r_{mn}f_{ba}^{n}f^{bm}_{c}+2r_{mn}f_{bc}^{n}f^{bm}_{a}=
% r_{an}B^{n}_{c}+r_{cn}B^{n}_{a} \eqno(3.31)
%$$
Hence
$$
 V= {1\ov 2}\d U  \eqno(3.29)
$$
The calculations of $U, V, W$ assembled together give
the result
$$
\eqalign{
 G^{0}(z)G^{1}(w)= (z-w)^{-2}({q\ov 2}r_{ab}-f_{ab}^{m}f^{n}r_{mn})
 \psi^{a}\psi^{b}+ \cr
 (z-w)^{-1}{1\ov 2}({q\ov 2}r_{ab}-f_{ab}^{m}f^{n}r_{mn})
 \d (\psi^{a}\psi^{b}) }
 \eqno(3.30)
$$
Taking into account the singular  terms of $G^{0}\d x^{1}$
and $\d x^{0}G^{1}$ the operator product (3.15) is given by
$$
\eqalign{
 G^{0}_{x}(z)G^{1}_{x}(w)= (z-w)^{-2}({q\ov 2}r_{ac}-f_{ac}^{m}f^{n}r_{mn}-
 {1\ov 2}x^{1}_{m}f_{ac}^{m}-
 {1\ov 2}x^{0}_{m}(r_{na}f^{nm}_{c}-
 r_{nc}f^{nm}_{a}))\psi^{a}\psi^{c} \cr
 +(z-w)^{-1}{1\ov 2}({q\ov 2}r_{ac}-f_{ac}^{m}f^{n}r_{mn}-
 x^{1}_{m}f_{ac}^{m})\d (\psi^{a}\psi^{c}) }\eqno(3.31)
$$
Next, we calculate
$$
 G^{0}_{x}(z)G_{1x}(w)=G^{0}(z)G_{1}(w)+G^{0}(z)\d x_{1}(w)
 +\d x^{0}(z)G_{1}(w)+\d x^{0}(z)\d x_{1}(w) \eqno(3.32)
$$
Here we start by calculating singular terms of $G^{0}G_{1}$
$$
\eqalign{
 G^{0}(z)G_{1}(w)=-(z-w)^{-3}{1\ov2}
 f_{am}^{n}f_{bc}^{d}r^{bm}r^{ca}r_{dn}+                         \cr
 (z-w)^{-2}\lbr
 (r^{ba}f_{ab}^{c}J_{c}-
 {1\ov2}r^{bc}<E_{a},E_{b}>\psi^{a}\psi_{c})                     \cr
 -{1\ov2}f_{am}^{n}f_{bc}^{d}r^{bm}r^{ca}r_{ds}\psi_{n}\psi^{s}+
 f_{am}^{n}f_{bc}^{d}r^{bm}r^{cq}r_{dn}\psi_{q}\psi^{a}\rbr+     \cr
 (z-w)^{-1}\lbr
 (r^{ba}f_{ab}^{c}\d J_{c}-
 {1\ov2}r^{bc}<E_{a},E_{b}>\d \psi^{a}\psi_{c})-                 \cr
 {1\ov2}f_{am}^{n}f_{bc}^{d}r^{bm}r^{ca}r_{ds}\d \psi_{n}\psi^{s}+
 f_{am}^{n}f_{bc}^{d}r^{bm}r^{cq}r_{dn}\psi_{q}\d \psi^{a}\rbr+  \cr
 (z-w)^{-1}\lbr
 f_{as}^{p}f_{bc}^{d}r_{bs}r^{cq}r_{dm}-{1\ov4}
 f_{am}^{n}f_{bc}^{d}r_{bp}r^{cq}r_{dn}\rbr
 \psi^{a}\psi^{m}\psi_{p}\psi_{q} }\eqno(3.33)
$$
Let us denote
$$
\eqalign{
 P= -{1\ov2}f_{am}^{n}f_{bc}^{d}r^{bm}r^{ca}r_{dn}                \cr
 Q= (r^{ba}f_{ab}^{c}J_{c}-
    {1\ov2}r^{bc}<E_{a},E_{b}>\psi^{a}\psi_{c})                    \cr
    -{1\ov2}f_{am}^{n}f_{bc}^{d}r^{bm}r^{ca}r_{ds}\psi_{n}\psi^{s}+
    f_{am}^{n}f_{bc}^{d}r^{bm}r^{cq}r_{dn}\psi_{q}\psi^{a}         \cr
 R= (r^{ba}f_{ab}^{c}\d J_{c}-
    {1\ov2}r^{bc}<E_{a},E_{b}>\d \psi^{a}\psi_{c})-                 \cr
    {1\ov2}f_{am}^{n}f_{bc}^{d}r^{bm}r^{ca}r_{ds}\d \psi_{n}\psi^{s}+
    f_{am}^{n}f_{bc}^{d}r^{bm}r^{cq}r_{dn}\psi_{q}\d \psi^{a}       \cr
 S= (f_{as}^{p}f_{bc}^{d}r_{bs}r^{cq}r_{dm}-{1\ov4}
    f_{am}^{n}f_{bc}^{d}r_{bp}r^{cq}r_{dn})
    \psi^{a}\psi^{m}\psi_{p}\psi_{q} }
$$
First we prove
$$
 S=0    \eqno(3.34)
$$
Let us denote
$$
 (r*f_{-})^{pq}_{n}= f_{bc}^{d}r^{bp}r^{cq}r_{dn} \eqno(3.35)
$$
and rewrite $S$ as follows
$$
\eqalign{
 {1\ov4}\lbr f_{as}^{p}(r*f_{-})^{sq}_{m}-
 f_{ms}^{p}(r*f_{-})^{sq}_{a}-             \cr
 f_{as}^{q}(r*f_{-})^{sp}_{m}+
 f_{ms}^{q}(r*f_{-})^{sp}_{a}-
 f_{am}^{n}(r*f_{-})^{pq}_{n}\rbr
 \psi^{a}\psi^{m}\psi_{p}\psi_{q} }\eqno(3.36)
$$
The right hand side of (3.36) is coboundary of 1-cochain
$r*f_{-}$ with coefficients in $\wedge^{2}g_{-}$.
{}From the other hand the cochain $r*f_{-}$ is coboundary
of 0- cochain $r$
$$
 (r*f_{-})^{pq}_{n}=r^{bp}f_{bn}^{q}-r^{bq}f_{bn}^{p} \eqno(3.37)
$$
Therefore equation (3.34) is the consequence of nilpotency
condition for the coboundary operator of Lie algebra $g_{-}$.
Because $r*f_{-}$ is coboundary of $r$ it defines bialgebra-
structure on $g_{-}$ [5]. Therefore we can do the change
$$
 f^{bm}_{d}\rightarrow (r*f_{-})^{bm}_{d}
$$
in the equations (2.2), (2.3). Using these new equations
one can get
$$
\eqalign{
 -{1\ov 2}f_{am}^{n}f_{bc}^{d}r^{bm}r^{ca}r_{dk}\psi_{n}\psi^{k}=
 {1\ov 2}f_{am}^{n}(r*f_{-})^{ma}_{k}\psi_{n}\psi^{k}=           \cr
 -{1\ov 2}(f_{m}(r*f_{-})^{mn}_{k}+ (r*f_{-})^{m}f_{mk}^{n}) }
 \eqno(3.38)
$$
{}From (3.18), (3,25) one can obtain
$$
\eqalign{
 f_{m}(r*f_{-})^{mn}_{s}=
 -{1\ov 2}r^{ab}f_{ab}^{i}f_{is}^{n} \cr
 (r*f_{-})^{m}f_{ms}^{n}=-{1\ov 2}r^{ab}f_{ab}^{m}f_{ms}^{n} }\eqno(3.39)
$$
Using (3.18) we can get
$$
 f_{am}^{n}f_{bc}^{d}r^{bm}r^{cq}r_{dn}\psi_{q}\psi^{n}=
 -(f_{m}r^{qm}f_{qk}^{n}+r^{pn}f_{pq}^{m}f_{mk}^{q})\psi_{n}\psi^{k}
 \eqno(3.40)
$$
Taking into account (3.38)-(3.40) one can write
$$
 Q= -2f_{a}r^{ab}(J_{b}+f_{bm}^{n}\psi_{n}\psi^{m}) \eqno(3.41)
$$
In the same way we rearrangement $R$
$$
 R= {1\ov 2}\d Q \eqno(3.42)
$$
Finaly for $P$ we obtain
$$
 P=0             \eqno(3.43)
$$
Indeed
$$
 P= {1\ov 2}f_{am}^{n}(r*f_{-})^{ma}_{n}= f_{m}r^{mn}f_{m}=0
$$
Summing up the resalts of culculations $P, Q, R, S$ we obtain
$$
\eqalign{
 G^{0}(z)G_{1}(w)= -(z-w)^{-2}2f_{a}r^{ac}(J_{c}+f_{cm}^{n}\psi_{n}\psi^{m})
 -(z-w)^{-1}f_{a}r^{ac}\d (J_{c}+f_{cm}^{n}\psi_{n}\psi^{m}) }
 \eqno(3.44)
$$
After calculation of the singular terms of $G^{0}\d x_{1}$,
$\d x^{0}G_{1}$, $\d x^{0}\d x_{1}$ we will obtain (3.14b)

The calculations of singular terms in operator products
$G_{0x}G_{1x}$ and $G_{0x}G^{1}_{x}$ are identical
with that just we have done.

 The proof is completed.

 To obtain $N=4$ Virasoro superalgebras operator products
one have to put either
$$
 G^{0}_{x}(z)G_{1x}(w)\sim G_{0x}(z)G^{1}_{x}(w)\sim 0 \eqno(3.45a)
$$
either
$$
 G_{0x}(z)G_{1x}(w)\sim G^{0}_{x}(z)G^{1}_{x}(w)\sim 0 \eqno(3.45b)
$$
Therefore there is two possibilities to construct generators of
$N=4$ Virasoro superalgebra. We will investigate each possibility.

 CASE (3.45a).
{}From (3.14b-c), (3.45a) one can obtain the system
of equations
$$
\eqalign{
 x^{0}_{a}x^{a}_{1}= 0                                           \cr
 x_{1}^{b}= f_{a}r^{ab}                                          \cr
 f_{a}r^{ab}(J_{b}+f_{bd}^{c}\psi_{c}\psi^{d})+
 x^{0}_{a}r^{ca}(J_{c}+f_{cb}^{d}r^{bn}r_{dk}\psi_{n}\psi^{k})=0 \cr
 x^{1}_{b}= f^{a}r_{ab}                                          \cr
 f^{a}r_{ab}(J^{b}+f^{bd}_{c}\psi^{c}\psi_{d})+
 x_{0}^{a}r_{ca}(J^{c}+f^{cb}_{d}r_{bn}r^{dk}\psi^{n}\psi_{k})=0 }
\eqno(3.46)
$$
Its solution is given by
$$
\eqalign{
 x_{1}^{b}= f_{a}r^{ab},\quad
 x_{0}^{a}=f^{a}                                                 \cr
 x^{1}_{b}= f^{a}r_{ab},\quad
 x^{0}_{a}=f_{a} }\eqno(3.47)
$$
Let us substitute the solution (3.47) into (3.14a)
$$
\eqalign{
 G^{0}_{x}(z)G^{1}_{x}(w)= (z-w)^{-2}\lbr
 {q\ov2}r_{ac}-{1\ov2}f_{ac}^{m}f^{n}r_{mn}
 -{1\ov2}f_{m}(r_{na}f^{nm}_{c}-r_{nc}f^{nm}_{a})
 \rbr \psi^{a}\psi^{c}+                                          \cr
 (z-w)^{-1}
 {q\ov4}r_{ac}\d (\psi^{a}\psi^{c}) }\eqno(3.48)
$$
{}From (2.2), (2.3), (3.18) it follows that
$$
 f_{m}(f^{mb}_{a}r_{bc}-f^{mb}_{c}r_{ba})=
 f_{ac}^{b}f^{m}r_{mb} \eqno(3.49)
$$
Indeed, from (2.2) one can get
$$
 f_{ac}^{k}f^{m}r_{mk}=
 {1\ov 2}f_{nm}^{k}(f^{nm}_{a}r_{kc}-f^{nm}_{c}r_{ka})
 \eqno(3.50)
$$
{}From the other hand, using (2.3), (3.18) one can get
$$
 f_{nm}^{k}(f^{nm}_{a}r_{kc}-f^{nm}_{c}r_{ka})=
 f_{m}(f^{mk}_{a}r_{kc}-f^{mk}_{c}r_{ka})-f^{m}f_{ac}^{k}r_{km}
 \eqno(3.51)
$$
Comparing (3.50) and (3.51) we obtain (3.49). Hence one may
write
$$
\eqalign{
 G^{0}_{x}(z)G^{1}_{x}(w)=
 (z-w)^{-2}{q\ov2}r_{ac}\psi^{a}\psi^{c}+
 (z-w)^{-1}
 {q\ov4}r_{ac}\d (\psi^{a}\psi^{c}) }\eqno(3.52)
$$
In the same way we can derive
$$
\eqalign{
 G_{0x}(z)G_{1x}(w)=
 (z-w)^{-2}{q\ov2}r^{ac}\psi_{a}\psi_{c}+
 (z-w)^{-1}
 {q\ov4}r^{ac}\d (\psi_{a}\psi_{c}) }\eqno(3.53)
$$
Motivated by formulas (3.52), (3.53) we redenote
currents $G^{a}_{x}(z)$, $G_{ax}(z)$, $a= 0, 1$ by
$$
 G_{ax}\rightarrow \sqrt{2\ov k+v}G_{ax},\quad
 G^{a}_{x}\rightarrow \sqrt{2\ov k+v}G^{a}_{x} \eqno(3.54)
$$
and introduce generators of $su(2)$- Kac-Moody algebra
$$
\eqalign{
 K^{01}= {1\ov2}r_{ac}\psi^{a}\psi^{c}, \cr
 K_{01}={1\ov2}r^{ac}\psi_{a}\psi_{c},  \cr
 K=\psi^{a}\psi_{a} }\eqno(3.55)
$$
Then (3.52), (3.53) shows that
$$
\eqalign{
 G^{0}_{x}(z)G^{1}_{x}(w)= (z-w)^{-2}4K^{01}(w)+
 (z-w)^{-1}2\d K^{01}(w)+ reg. \cr
 G_{0x}(z)G_{1x}(w)= (z-w)^{-2}4K_{01}(w)+
 (z-w)^{-1}2\d K_{01}(w)+ reg. }\eqno(3.56)
$$
As a simple exercise in the application of formulas
(3.10), (3.18), (3.24) one may obtain
$$
\eqalign{
 K(z)G^{0}_{x}(w)= (z-w)^{-1}G^{0}_{x}(w)+ reg. \cr
 K(z)G_{0x}(w)= -(z-w)^{-1}G_{0x}(w)+ reg. \cr
 K^{01}(z)G^{0}_{x}(w)=0 \cr
 K_{01}(z)G_{0x}(w)=0 \cr
 K_{01}(z)G^{0}_{x}(w)= -(z-w)^{-1}G_{1x}(w)+reg. \cr
 K^{01}(z)G_{0x}(w)= -(z-w)^{-1}G^{1}_{x}(w)+reg. }\eqno(3.57)
$$
To find the stress-energy tensor $T$ we calculate operator
product $G^{0}_{x}G_{0x}$, but the result follows from
the Manin triple construction for $N=2$ Virasoro
superalgebra (2.9)
$$
\eqalign{
 T={1\over 2(k+v)}(J^{a}J_{a}+J_{a}J^{a})+
 (\d \psi^{a}\psi_{a}-\psi^{a}\d \psi_{a})+                      \cr
 {1\over 2(k+v)}\d (f_{a}J^{a}-f^{a}J_{a})+
 {1\over 2(k+v)}(f_{a}f^{ab}_{c}-f^{a}f_{ac}^{b})
 \d (\psi^{c}\psi_{b}) }\eqno(3.58)
$$
Taking into account (3.49) it is not dificult to
show that currents (3.55) are dimension one primary
fields relative stress-tensor (3.58).
It is clear that OPE $G^{1}_{x}G_{1x}$ gives us the
same stress-energy tensor (3.62) because currents
$G^{1}_{x}(z), G_{1x}(z)$ can be derived from
currents $G^{0}_{x}(z), G_{0x}(z)$ with help of
transformation $\psi^{a}\rightarrow r_{ab}\psi^{b}$,
$\psi_{a}\rightarrow r^{ab}\psi_{b}$.
The following lemma sums up our investigation
of CASE (3.45a)

LEMMA 3.2.
 The fermionic currents
$$
\eqalign{
 G^{0}= \sqrt{2\ov k+v}(\psi^{a}J_{a}-
 {1\over 2}f_{ab}^{c}\psi^{a}\psi^{b}\psi_{c}
 +f_{a}\d \psi^{a})                                              \cr
 G_{0}= \sqrt{2\ov k+v}(\psi_{a}J^{a}-
 {1\over 2}f^{ab}_{c}\psi_{a}\psi_{b}\psi^{c}
 +f^{a}\d \psi_{a})                                              \cr
 G^{1}_{x}= \sqrt{2\ov k+v}(r_{ba}\psi^{a}J^{b}+
 {1\over 2}r_{am}f^{ab}_{c}r_{bn}r^{ck}\psi^{m}\psi^{n}\psi_{k}
 +f^{a}r_{ab}\d \psi^{b})                                        \cr
 G_{1x}= \sqrt{2\ov k+v}(r^{ba}\psi_{a}J_{b}+
 {1\over 2}r^{am}f_{ab}^{c}r^{bn}r_{ck}\psi_{m}\psi_{n}\psi^{k}
 +f_{a}r^{ab}\d \psi_{b}) }\eqno(3.59)
$$
generate $N=4$ Virasoro superalgebra with central charge
$$
 c= 3d  \eqno(3.60)
$$
stress- energy tensor
$$
\eqalign{
 T={1\over 2(k+v)}(J^{a}J_{a}+J_{a}J^{a})+
 (\d \psi^{a}\psi_{a}-\psi^{a}\d \psi_{a})+                      \cr
 {1\over 2(k+v)}\d (f_{a}J^{a}-f^{a}J_{a})+
 {1\over 2(k+v)}(f_{a}f^{ab}_{c}-f^{a}f_{ac}^{b})
 \d (\psi^{c}\psi_{b}) }
 \eqno(3.61)
$$
and $su(2)$- Kac-Moody currents
$$
\eqalign{
 K^{01}= {1\ov2}r_{ac}\psi^{a}\psi^{c} \cr
 K_{01}={1\ov2}r^{ac}\psi_{a}\psi_{c}  \cr
 K=\psi^{a}\psi_{a} }
 \eqno(3.62)
$$

 CASE (3.45b).
{}From (3.14a), (3.14d), (3.45b) we obtain the following equations
$$
\eqalign{
 {q\ov2}r_{ac}-f_{ac}^{m}f^{n}r_{nm}-x^{1}_{m}f_{ac}^{m}=0       \cr
 {q\ov2}r_{ac}-f_{ac}^{m}f^{n}r_{nm}-
 x^{0}_{m}(r_{na}f^{nm}_{c}-r_{nc}f^{nm}_{a})=0                  \cr
 {q\ov2}r^{ac}-f^{ac}_{m}f_{n}r^{nm}-x_{1}^{m}f^{ac}_{m}=0       \cr
 {q\ov2}r^{ac}-f^{ac}_{m}f_{n}r^{nm}-
 x_{0}^{m}(r^{na}f_{nm}^{c}-r^{nc}f_{nm}^{a})=0 }
 \eqno(3.63)
$$
The first and third equations of this system have the
solutions if nondegenerate 2-cocycles $r, r^{-1}$
are the coboundary cocycles:
$$
 r_{ac}=r_{m}f_{ac}^{m},\quad
 r^{ac}=r^{m}f^{ac}_{m}
 \eqno(3.64)
$$
In this case the solutions of the first and third
equations are given by
$$
\eqalign{
 x^{1}_{m}=-r_{mn}f^{n}+{q\ov2}r_{m}  \cr
 x_{1}^{m}=-r^{mn}f_{n}+{q\ov2}r^{m} }
 \eqno(3.65)
$$
Solving remain equations we find the conditions
such that (3.49b) is satisfied
$$
\eqalign{
 r_{ac}=r_{m}f_{ac}^{m},\quad
 r^{ac}=r^{m}f^{ac}_{m}                                          \cr
 x^{1}_{m}=-r_{mn}f^{n}+{q\ov2}r_{m},\quad
 x_{1}^{m}=-r^{mn}f_{n}+{q\ov2}r^{m}                             \cr
 x^{0}_{m}=f_{m}+{q\ov2}r^{n}r_{nm},\quad
 x_{0}^{m}=f^{m}+{q\ov2}r_{n}r^{nm} }
 \eqno(3.66)
$$
Let us substitute the solutions (3.66) into (3.14b):
$$
\eqalign{
 G^{0}_{x}(z)G_{1x}(w)= (z-w)^{-2}qr^{b}
 (J_{b}+f_{bd}^{c}\psi_{c}\psi^{d})+
 (z-w)^{-1}{q\ov2}r^{b}\d (J_{b}+f_{bd}^{c}\psi_{c}\psi^{d})     \cr
 G_{0x}(z)G^{1}_{x}(w)= (z-w)^{-2}qr_{b}
 (J^{b}+f^{bd}_{c}\psi^{c}\psi_{d})+
 (z-w)^{-1}{q\ov2}r_{b}\d (J^{b}+f^{bd}_{c}\psi^{c}\psi_{d}) }
 \eqno(3.67)
$$
Motivated by these formulas we redenote currents
$G^{a}_{x}(z)$, $G_{ax}(z)$, $a= 0, 1$ by (3.54)
and introduce $su(2)$-Kac-Moody currents:
$$
\eqalign{
 K^{0}_{1}= r^{a}(J_{a}+f_{ab}^{c}\psi_{c}\psi^{b})              \cr
 K^{1}_{0}= r_{a}(J^{a}+f^{ab}_{c}\psi^{c}\psi_{b})              \cr
 K=(\delta^{b}_{a}-r_{c}r^{cd}f_{da}^{b}-
 r^{c}r_{cd}f^{da}_{b})\psi^{a}\psi_{a}+
 r_{b}r^{ba}J_{a}-r^{b}r_{ba}J^{a} }
 \eqno(3.68)
$$
The stress-energy tensor may be obtained in a
similar way to the CASE (3.45a):
$$
\eqalign{
 T={1\over 2(k+v)}(J^{a}J_{a}+J_{a}J^{a})+
 (\d \psi^{a}\psi_{a}-\psi^{a}\d \psi_{a})+                    \cr
 {1\over 2(k+v)}\d ((f_{a}+{q\ov2}r^{b}r_{ba})J^{a}-
 (f^{a}+{q\ov2}r_{b}r^{ba})J_{a})+                             \cr
 {1\over 2(k+v)}((f_{a}+{q\ov2}r^{b}r_{ba})f^{ab}_{c}-
 (f^{a}+{q\ov2}r_{b}r^{ba})f_{ac}^{b})
 \d (\psi^{c}\psi_{b}) }
 \eqno(3.69)
$$
Summing up the investigation of CASE (3.45b) we
can get the following

LEMMA 3.3
 If nondegenerate 2-cocycles are coboundary
$$
 r_{ac}=r_{m}f_{ac}^{m},\quad
 r^{ac}=r^{m}f^{ac}_{m}
 \eqno(3.70)
$$
then fermionic currents
$$
\eqalign{
 G^{0}= \sqrt{2\ov k+v}(\psi^{a}J_{a}-
 {1\over 2}f_{ab}^{c}\psi^{a}\psi^{b}\psi_{c}
 +(f_{a}+{q\ov2}r^{b}r_{ba})\d \psi^{a})                         \cr
 G_{0}= \sqrt{2\ov k+v}(\psi_{a}J^{a}-
 {1\over 2}f^{ab}_{c}\psi_{a}\psi_{b}\psi^{c}
 +(f^{a}+{q\ov2}r_{b}r^{ba})\d \psi_{a})                         \cr
 G^{1}_{x}= \sqrt{2\ov k+v}(r_{ba}\psi^{a}J^{b}+
 {1\over 2}r_{am}f_{ab}^{c}r_{bn}r^{ck}\psi^{m}\psi^{n}\psi_{k}
 +({q\ov2}r_{a}-r_{ab}f^{b})\d \psi^{a})                         \cr
 G_{1x}= \sqrt{2\ov k+v}(r^{ba}\psi_{a}J_{b}+
 {1\over 2}r^{am}f^{ab}_{c}r^{bn}r_{ck}\psi_{m}\psi_{n}\psi^{k}
 +({q\ov2}r^{a}-r^{ab}f_{b})\d \psi_{a}) }
 \eqno(3.71)
$$
generate $N=4$ Virasoro superalgebra with central charge
$$
 c= 3(qr^{a}r_{a}-d) \eqno(3.72)
$$
stress- energy tensor
$$
\eqalign{
 T={1\over 2(k+v)}(J^{a}J_{a}+J_{a}J^{a})+
 (\d \psi^{a}\psi_{a}-\psi^{a}\d \psi_{a})+                    \cr
 {1\over 2(k+v)}\d ((f_{a}+{q\ov2}r^{b}r_{ba})J^{a}-
 (f^{a}+{q\ov2}r_{b}r^{ba})J_{a})+                             \cr
 {1\over 2(k+v)}((f_{a}+{q\ov2}r^{b}r_{ba})f^{ab}_{c}-
 (f^{a}+{q\ov2}r_{b}r^{ba})f_{ac}^{b})
 \d (\psi^{c}\psi_{b}) }
 \eqno(3.73)
$$
and $su(2)$-Kac-Moody currents
$$
\eqalign{
 K^{0}_{1}= r^{a}(J_{a}+f_{ab}^{c}\psi_{c}\psi^{b})              \cr
 K^{1}_{0}= r_{a}(J^{a}+f^{ab}_{c}\psi^{c}\psi_{b})              \cr
 K=(\delta^{b}_{a}-r_{c}r^{cd}f_{da}^{b}-
 r^{c}r_{cd}f^{da}_{b})\psi^{a}\psi_{a}+
 r_{b}r^{ba}J_{a}-r^{b}r_{ba}J^{a} }
 \eqno(3.74)
$$

 Now we are in a position to formulate the conditions
such that N=2 SCFT associated with any finite-
dimensional Manin triple possess $N=4$ Virasoro
superalgebra of symmetries. To do it let us
introduce Drinfeld's definition of quasi Frobenius
and Frobenius Lie algebras:

DEFINITION 3.4. [6]
 Finite- dimensional Lie algebra is called quasi Frobenius
Lie algebra if it endowed with nondegenerate 2-cocycle.
If its cocycle is coboundary then it is called Frobenius
Lie algebra.

 Due to this definition we will call quasi Frobenius
Manin triple a Manin triple with quasi Frobenius
isotropic subalgebras such that the corresponding
nondegenerate 2-co\-cyc\-les are mutualy inverse.
If they are coboundary cocycles we will call
this Manin triple Frobenius Manin triple.

 As a consequence of lemmas 3.2, 3.3 we can get

PROPOSITION 3.5
 Any $N=2$ SCFT associated with quasi Frobenius
Manin triple admits $N=4$ extention by the formulas
(3.59)- (3.62). If a Manin triple is Frobenius
Manin triple then $N=2$ SCFT admits two $N=4$
extentions by the formulas (3.59)- (3.62)
and (3.70)- (3.74).

 Let us make contact with paper [3] where "big"
$N=4$ Virasoro superalgebra was constructed.
{}From the formulas of [3] one can observe that the
modification (in the notations used in [3])
$$
\eqalign{
 T(z)\rightarrow \hat{T}(z)= T(z)+(1-\gamma)\d U(z) \cr
 G_{a}(z)\rightarrow \hat{G}_{a}(z)=G_{a}(z)+
 2(1-\gamma)\d \Gamma_{a}(z) }
 \eqno(3.75)
$$
converts "big" $N=4$ Virasoro superalgebra into
usual $N=4$ Virasoro superalgebra with generators
(3.70)-(3.74). The modification
$$
\eqalign{
 T(z)\rightarrow \hat{T}(z)= T(z)+\gamma \d U(z) \cr
 G_{a}(z)\rightarrow \hat{G}_{a}(z)=G_{a}(z)+
 2\gamma \d \Gamma_{a}(z) }
 \eqno(3.76)
$$
converts "big" $N=4$ Virasoro superalgebra into
usual $N=4$ Virasoro superalgebra with generators
(3.59)- (3.62). Therefore construction of "big"
$N=4$ Virasoro superalgebra is possible only for
the Frobenius Manin triple.

%%%%%%%%%%%%%%%%%%%%%%%%%%%%%%%%%%%%%%%%%%%%%%%%%%%%%%%%%%%%%%%%%%%%%

\vskip 10pt
\centerline{\bf4. Examples.}
%%%%%%%%%%%%%%%%%%%%%%%%%%%%%%%%%%%%%%%%%%%%%%%%%%%%%%%%%%%%%%%%%%%%%%

 EXAMPLE 1.
  The first example of the Manin triple
bases on any simple Lie algebra
$g$ with the scalar product $(,)$ and its Cartan
decomposition $g=n_{-}\oplus h\oplus n_{+}$, $b_{+}=
h\oplus n_{+}$, $b_{-}=h\oplus n_{-}$.
Consider the Lie algebra
$$
      p=g\oplus \tilde{h} \eqno(4.1)
$$
where $\tilde{h}$ is the copy of the
Cartan subalgebra $h$ and the Lie algebra
structure on $p$ is defined by
$$
       [g,\tilde{h}]=0 \eqno(4.2)
$$
On $p$ we define the invariant scalar product
$$
     <(X_{1},H_{1}),(X_{2},H_{2})>=(X_{1},X_{2})-(H_{1},H_{2}) \eqno(4.3)
$$
If we set
$$
\eqalign { p_{+}=\{(X,H)\in p|X\in b_{+},\quad H=X_{h}\} \cr
           p_{-}=\{(X,H)\in p|X\in b_{-},\quad H=-X_{h}\} } \eqno(4.4)
$$
,where $X_{h}$ is the projection of $X$
on the Cartan subalgebra $h$,
then we will have $p=p_{+}\oplus p_{-}$ and
$p_{+}$, $p_{-}$ are isotropic subalgebras of $p$,
wich are isomorphic to Borel subalgebras $b_{+}$, $b_{-}$.
We will give the explicit $N=4$ Virasoro superalgebra
construction in the simplest case
$$
 g=sl(2,C)   \eqno(4.5)
$$
In this case there is only one way to fix
nondegenerate 2- cocycles on isotropic subalgebras $p_{\pm}$,
namely in the orthonormal basis (2.1) they are given by
$$
\eqalign{
 r(E_{0},E_{1})=r_{01}=-r^{-1} \cr
 r^{-1}(E^{0},E^{1})=r^{01}=r }\eqno(4.6)
$$
, where r is arbitrary nonzero complex number.
Cocycles (4.6) are coboundary cocycles
$$
 r_{01}=-r^{-1}f_{01}^{1},\quad r^{01}=-rf^{01}_{1} \eqno(4.7)
$$
Therefore formulas (4.1), (4.4)- (4.6) define
Frobenius Manin triple.
In this case let $J^{a},J_{a},\psi^{a},\psi_{a}$, $a=0,1$
be the bosonic and fermionic currents
with the OPE
$$
\eqalign { J^{0}(z)J^{1}(0)=-z^{-1}J^{1}(0)+o(z) \cr
           J_{0}(z)J_{1}(0)=z^{-1}J_{1}(0)+o(z) \cr
           J^{0}(z)J^{0}(0)=-z^{-2}+o(z)        \cr
           J_{0}(z)J_{0}(0)=-z^{-2}+o(z)        \cr
           J^{0}(z)J_{1}(0)=z^{-1}J_{1}(0)+o(z) \cr
           J^{1}(z)J_{0}(0)=z^{-1}J^{1}(0)+o(z)  \cr
           J^{0}(z)J_{0}(0)=z^{-2}(k+1)+o(z)     \cr
           J^{1}(z)J_{1}(0)=z^{-2}k-z^{-1}(J_{0}+J^{0})(0)+o(z) \cr
           \psi^{a}(z)\psi_{b}(0)=z^{-1}\delta^{a}_{b}+o(z) }\eqno(4.8)
$$
Then the formulas (3.59)- (3.62) have the
following form
$$
\eqalign{
 G^{0}= \sqrt{2\ov k+2}(\psi^{0}J_{0}+\psi^{1}J_{1}-
 \psi^{0}\psi^{1}\psi_{1}+\d \psi^{0}) \cr
 G_{0}= \sqrt{2\ov k+2}(\psi_{0}J^{0}+\psi_{1}J^{1}+
 \psi_{0}\psi_{1}\psi^{1}-\d \psi_{0}) \cr
 G^{1}= -\sqrt{2\ov k+2}r^{-1}(\psi^{1}J^{0}-\psi^{0}J^{1}+
 \psi^{1}\psi^{0}\psi_{0}-\d \psi^{1}) \cr
 G_{1}= \sqrt{2\ov k+2}r(\psi_{1}J_{0}-\psi_{1}J_{0}+
 \psi_{1}\psi_{0}\psi^{0}-\d \psi_{1}) }
 \eqno(4.9)
$$
$$
 c=6 \eqno(4.10)
$$
$$
\eqalign{
 T= {1\ov 2(k+2)}(J_{a}J^{a}+J^{a}J_{a})+
 {1\ov 2}(\d \psi^{a}\psi_{a}-\psi^{a}\d \psi_{a})+
 {1\ov 2(k+2)}\d (J^{0}+J_{0}) \cr
 K^{01}= -r^{-1}\psi^{0}\psi^{1},\quad
 K= \psi^{a}\psi_{a},\quad
 K_{01}= r\psi_{0}\psi_{1} }
 \eqno(4.11)
$$
For the second $N=4$ Virasoro superalgebra
formulas (3.70)- (3.74) will look like
$$
\eqalign{
 G^{0}= \sqrt{2\ov k+2}(\psi^{0}J_{0}+\psi^{1}J_{1}-
 \psi^{0}\psi^{1}\psi_{1}-(k+1)\d \psi^{0}) \cr
 G_{0}= \sqrt{2\ov k+2}(\psi_{0}J^{0}+\psi_{1}J^{1}+
 \psi_{0}\psi_{1}\psi^{1}+(k+1)\d \psi_{0}) \cr
 G^{1}= -\sqrt{2\ov k+2}r^{-1}(\psi^{1}J^{0}-\psi^{0}J^{1}+
 \psi^{1}\psi^{0}\psi_{0}+(k+1)\d \psi^{1}) \cr
 G_{1}= \sqrt{2\ov k+2}r(\psi_{1}J_{0}-\psi_{1}J_{0}+
 \psi_{1}\psi_{0}\psi^{0}+(k+1)\d \psi_{1}) }
 \eqno(4.12)
$$
$$
 c=6(k+1) \eqno(4.13)
$$
$$
\eqalign{
 T= {1\ov 2(k+2)}(J_{a}J^{a}+J^{a}J_{a})+
 {1\ov 2}(\d \psi^{a}\psi_{a}-\psi^{a}\d \psi_{a})+
 {(k+1)\ov 2(k+2)}\d (-J^{0}+J_{0}) \cr
 K^{0}_{1}= -r(J_{1}-\psi_{1}\psi^{0}) \cr
 K= \psi^{0}\psi_{0}-\psi^{1}\psi_{1}+J_{0}+J^{0} \cr
 K^{1}_{0}= -r^{-1}(J^{1}+\psi^{1}\psi_{0}) }
 \eqno(4.14)
$$
This construction was used in [10] to prove
N=4 Virasoro superalgebras determinant formula
[11-12].

 EXAMPLE 2.
Let $g$ be simple even- dimensional Lie algebra.
In this situation we can represent its Cartan
subalgebra $h$ as the direct sum of subspaces
isotropic with respect to the Killing form :
$$
    h=h_{+}\oplus h_{-} \eqno(4.15)
$$
If we set
$$
\eqalign{ p_{+}=n_{+}\oplus h_{+} \cr
          p_{-}=n_{-}\oplus h_{-} }\eqno(4.16)
$$
then we will have $g=p_{+}\oplus p_{-}$ and $p_{+}$, $p_{-}$
are isotropic subalgebras of $g$.
  In this example we give the explicit N=4 Virasoro
superalgebra construction  in the case:
$$
    g=sl(3,C) \eqno(4.17)
$$
Let $\lbrc E_{3}, E_{2}, E_{1}$,$H_{1}, H_{2}$,
$E^{1}, E^{2}, E^{3}\rbrc$
be the standard basis in $sl(3,C)$, such that generators
$E^{1}, E^{2}$ correspond to the simple roots
$\alpha_{1}, \alpha_{2}$, $E^{3}$ corresponds to the maximal
root $\alpha_{3}$ and $E_{1},...,E_{3}$ correspond to the
negative roots of $sl(3,C)$. We define invariant inner product
$(,)$ on $g$ by the formula
$$
 (x,y)= Tr(xy) \eqno(4.18)
$$
, where $x, y \in g$ and are $3\times 3$ matrixes.
The basises in isotropic subalgebras $p_{\pm}$
constituting orthonormal basis in $g$ are given by
$$
 p_{+}=\oplus_{a=0}^{3}CE^{a},\quad
 p_{-}=\oplus_{a=0}^{3}CE_{a} \eqno(4.19)
$$
, where
$$
 E^{0}= {1\ov \sqrt{3}}(H_{1}+exp(\imath {\pi \ov 3})H_{2}),\quad
 E_{0}= {1\ov \sqrt{3}}(H_{1}+exp(-\imath {\pi \ov 3})H_{2})
 \eqno(4.20)
$$
Next, one need to fix nondegenerate 2-cocycles on
isotropic subalgebras. By the direct culculation
one can find that the skew- symmetric bilinear form
r is cocycle on $p_{+}$ if the following equations
are satisfied
$$
 r^{12}= {r^{03}\ov \alpha_{3}(E^{0})},\quad
 r^{13}=r^{23}=0 \eqno(4.21)
$$
, where $r^{ab}=r(E^{a},E^{b})$. Cocycle r is nondegenerate
if $r^{03}$ is nonzero. From (4.21) it follows that r is coboundary
cocycle
$$
\eqalign{
 r^{ab}= r^{c}f^{ab}_{c} \cr
 r^{a}= {r^{0a}\ov \alpha_{a}(E^{0})} }
 \eqno(4.22)
$$
The same is true for nondegenerate 2- cocycles on
subalgebra $p_{-}$. That is if we put
$$
 r_{a}= {1\ov \alpha_{a}(E_{0})r^{0a}} \eqno(4.23)
$$
we obtain nondegenerate coboundary 2- cocycle $r^{-1}$
on $p_{-}$
$$
 r_{ab}= r_{c}f_{ab}^{c} \eqno(4.24)
$$
which is invers to the 2- cocycle r on $p_{+}$.
Hence one may conclude the formulas (4.17)- (4.24)
define Frobenius Manin triple and (3.63)-(3.66),
(3.74)- (3.78) give us two N=4 Virasoro superalgebras.

 EXAMPLE 3.
We construct N=4 Virasoro superalgebra based on quasi
Frobenius Manin triple with 4-dimensional nilpotent
isotropic subalgebras. Let $g_{+}$ be 4- dimensional
nondecomposible nilpotent Lie algebra [6].
There is the only one Lie algebra of this type:
$$
\eqalign{
 g_{+}=\oplus_{a=1}^{4}CE^{a} \cr
 [E^{1}, E^{2}]=E^{3},\quad [E^{1}, E^{3}]=E^{4} }
 \eqno(4.25)
$$
(other brakets are equal to zero).
By the direct calculations one obtain that the skew-
symmetric bilinear form r on $g_{+}$ is cocycle if
$$
 r^{24}= r^{34}= 0 \eqno(4.26)
$$
and cocycle r is nondegenerate if
$$
 r^{14}\neq 0, r^{23}\neq 0 \eqno(4.27)
$$
Fathermore any
2-cocycle r is coboundary iff
$$
 r^{14}=r^{23}=0 \eqno(4.28)
$$
{}From (4.26)- (4.28) it follows that
if the equations (4.26), (4.27) are satisfied
then $g_{+}$ will be quasi Frobenius Lie algebra.
For simplicity we set
$$
 r^{12}= r^{13}= 0 \eqno(4.29)
$$
Then the invers matrix $r^{-1}$ have the following
nonzero elements
$$
\eqalign{
 r_{14}= -r_{41}= {1\ov r^{14}} \cr
 r_{23}= -r_{32}= {1\ov r^{23}} }
 \eqno(4.30)
$$
Having $r^{-1}\in \wedge^{2}g_{+}$ one can use it to define
the coboundary bialgebra structure on $g_{+}$ [5].
Let $g_{-}$ be the dual space to $g_{+}$ and
$E_{1},...,E_{4}$ be the dual basis to the basis (4.25)
$$
\eqalign{
 g_{-}= \oplus_{a=1}^{4}CE_{a},\quad
 (E_{a}, E^{b})=\delta^{b}_{a} }
 \eqno(4.31)
$$
then the Lie algebra structure on $g_{-}$ defined
by coboundary cocommutator on $g_{+}$ is given by
$$
 [E_{4}, E_{2}]= {1\ov r^{23}}E_{1},\quad
 [E_{4}, E_{1}]= -{1\ov r^{14}}E_{2} \eqno(4.32)
$$
In view of one- to- one correspondence between
Lie bialgebras and Manin triples [5] we obtain
the Manin triple $(g, g_{+}, g_{-})$.
Moreover $g_{-}$ is also quasi Frobebius Lie algebra
because as it follows from (4.32) $r^{-1}$
defines the isomorphism of Lie algebras
$$
 r^{-1}: g_{-}\rightarrow g_{+}
$$
such that preimage of the cocycle $r$ is equal to $r^{-1}$.
Therefore we conclude that the formulas (4.25)- (4.32)
define quasi Frobenius Manin triple, and formulas (3.59)-
(3.62) give us N=4 Virasoro superalgebra.
%%%%%%%%%%%%%%%%%%%%%%%%%%%%%%%%%%%%%%%%%%%%%%%%%%%%%%%%%%%%%%%%%%%%%%%%%%%%

\vskip 10pt
\centerline{\bf REFERENCES}
\frenchspacing
\item{[1]}Y.Kazama, H.Suzuki, {\it  Mod.Phys.Lett}
\ {\bf  A4}\ (1989)\ 235;
\ {\it Phys.Lett.}\ {\bf 216B}\ (1989)\ 112;
\ {\it Nucl.Phys.}\ {\bf B321}\ (1989)\ 232.
\item{[2]}P.Spindel, A.Sevrin, W.Troost, A.Van Proeyen, {\it  Nucl.Phys}
\ {\bf  B308}\ (1988)\ 662;
\ {\bf B311}\ (1989/89)\ 465.
\item{[3]} S.Parkhomenko, {\it Zh. Eksp. Teor. Fiz.}
\ {\bf 102}\ (July 1992)\ 3-7.
\item{[4]}E.Getzler, Manin Pairs and topological Field Theory,
\ {\it  MIT-preprint}\ (??? 1994).
\item{[5]}V.G.Drinfeld, Quantum groups,
\ {\it  Proc. Int. Cong. Math., Berkley, Calif.}\ (1986)\ 798.
\item{[6]}A.G.Elashvili, Frobenius Lie algebras 2,
\ {\it  Works of Tbilissi Math. Institute}
\ {\bf v.LXXVII}\ (1985)\ ???.
\item{[7]}M.Gunaydin, J.L.Petersen, A.Taormina, A.Van Proeyen,
{\it Nucl.Phys.}\ {\bf B322}\ (1989)\ 402.
\item{[8]}J.L.Petersen, A.Taormina, {\it CERN-TH.5446/89.};
\ {\it EFI-90-61}\ (August 1990);
\ {\it CERN-TH.5503/89}.
\item{[9]}H.Ooguri, J.L.Petersen, A.Taormina,
{\it Nucl. Phys.}\ {\bf B368}\ (1992)\ 611.
\item{[10]}S.Matsuda, {\it  Phys.Lett.}\ {\bf 282B}\ (1992)\ 56;
\item{[11]}W.Boucher,D.Friedan,A.Kent, {\it  Phys.Lett.}
\ {\bf 172B}\ (1986)\ 316;
\item{[12]}V.K.Dobrev, {\it  Phys.Lett.}\ {\bf 186B}\ (1987)\ 43;
\item{[13]}D.I.Gurevich, V.V.Lychagin, V.N.Rubtsov,
Nonholonomic filtration of cohomologies of Lie algebras
and "Large brackets", {\it Translated Matematicheskie Zametki}
\ {\bf Vol.52, No.1}\ (1992)\ 36.

\vfill\eject
%\end{document}
\end